# Narrow band emission from Au-film on Cu- substrate.

*Kompan M.E., Melekh B.A.-T., Nikitin S.E.*
A.F. Ioffe Institute , Saint Petersburg, Russia

**Abstract**

The experiments reveal the strong narrow band in the spectra of the emission from Au-film on the Cu substrate. The emission was excited by red He-Ne laser in standard geometry for Raman back-scattering experiments. Some additional experiments were made. The set of obtained data did not give enough arguments to make the final conclusion about the nature of this emission. The details of experiments, results, and possible explanations are discussed.

**Samples and the experimental techniques.**

The experiments were held in standard Raman back-scattering geometry. Red light from He-Ne laser (6328A; 1-5 mW) normally falls on surface of sample under investigations. Emission from the sample was collected by microscope Olimpus BX42, mainly with long-focus objective x20, and analyzed by micro-Raman spectrograph Horiba Jobin-Yvon MRS 320 with CCD detector. Experiments were done at room temperature.

Samples under investigation were gold films on copper substrates. Also samples with intermediate layer $MoS_2$ were studied in test experiments. The interlayer was made by mechanical pilling of $MoS_2$ preliminary to Au film formation.

Gold films were made by magnetron sputtering. Parameters of magnetron sputtering were as follows: total pressure was 1.3 Pa; partial pressure of oxygen was about 0.2 Pa; rate of gold deposition – 8-10 nm·min$^{-1}$. Thickness of gold film varied from 10 to 20 nm. Copper for substrates was pure thick cupper foil (99,999%) for electronic industry. Initially copper plate was covered by plastic film to save it from pollutions. The covering film was taken off just before Au sputtering. The relief of copper was not too smooth - the optical and AFM inspection revealed the traces of metallurgical treatment with the height about $\mu$m.

**Results of the observations. Band near 2150 cm$^{-1}$.**

The most interesting fact was the observation of the strong narrow band in spectra of the emission from the sample. As far as our installation did not allow us to distinguish whether it was due to inelastic scattering or due to re-emission (luminescence) we use in this paper versatile term "emission". Later in paper we shall discuss the possibility to attribute this band. The position of the band was about 2150 cm$^{-1}$ to Stocks side from exciting light 633nm, or 733nm in absolute units.

There were other bands in spectra, and the positions and even existence of those additional bands depended on the sample and on the conditions of experiment.

Several examples of spectra are at fig 1,2,. These and the other spectra in this paper were not processed (no baseline subtracting, ets.) in order not to omit the details.

At every spectrum (fig.1,2) the most notable is the strong band near 2150 cm$^{-1}$. The exact position could vary from sample to sample up to ±5 cm$^{-1}$. Sometimes it was evident by eye that the band consists from subbands (e.g. fig.2), but mostly it was smooth, strong and narrow.

**Results of the observations. Band in region 220 - 550 cm$^{-1}$.**

Apart from most powerful band near 2150 cm$^{-1}$ the spectra contain the specific double bands in interval 220 - 550 cm$^{-1}$. Notable, that band near 500-550 cm$^{-1}$ can be seen in spectrum of cooper substrate (fig.3), and in case, when the light was focused at very edge of Au film, at place, where the amount of gold is minor (fig.2). In case, then copper is essentially covered by Au (fig.1) the position of band is quite another - 220 cm$^{-1}$. At the same time neither Cu substrate nor the gold film itself do not demonstrate similar band (fig.3). We suppose, that this band is due light scattering at localized plasmons, and the shift of it's position is due changes at interface. The analyzes of this last observation will be the subject of communication elsewhere.

**Additional Raman experiments.**

Optical emission from metallic sample is not common-observed phenomenon. So the first set of additional experiments had the aim to cancel the ability of experimental mistakes. The spectra of light scattered from opaque transparent media did not show any extra lines in laser light, but 633 nm, used for excitation.

Second set of supplementary experiments include the record of spectra from components of the sample separately. The spectra were recorded from plastic cover film; from Cu substrate itself; also from gold films, sputtered on glass and on silicon. 3. Spectra from those components are shown at fig. 3. No one of those spectra contain the band near 2150 $cm^{-1}$ . At the same time the line of scattering in Si is well seen in spectrum. That's why one can conclude, that observed emission band 2150 $cm^{-1}$ is really originates from Au film on Cu substrate.

Also necessary to note, that double band near 550 $cm^{-1}$ is present in Cu-substrate spectra.

The attempt was made to influence someway on the sample. The Au-Cu sample was heated for 30 minutes at 250 C in flow of pure nitrogen . The band 2150 $cm^{-1}$ was observed before this thermal treatment, but was not observed after it.  We shall discus this fact later.

The experiments were done with samples with intermediate $MoS_2$-layer. Spectra of emission from those samples are shown at figs. 4,5. One can easy observe additional band near 1590 $cm^{-1}$ and the multiple bands from $MoS_2$ around 500 $cm^{-1}$ (for comparison: spectra from $MoS_2$ [1]). The idea and discussion of this experiment will be in paper later.

In order to get additional arguments to understand the nature of emission, the gold was sputtered on smooth Cu sample. The band 2150 $cm^{-1}$ was not observed in this case.

Also the attempt was made to distinguish whether the band under consideration originates from the scattering or from the luminescence. In this case the 532 nm laser was used for excitation. Unexpectedly, no correspond band was observe neither near position 2150 $cm^{-1}$ from 532 nm,  nor at 733 nm . Nevertheless, in case of sample with intermediate $MoS_2$ layer, with 532nm excitation, there were observed the band near 1590 $cm^{-1}$ and the bands near 500-550 $cm^{-1}$.

**Discussion of the results.**

There are no enough data to clarify the nature of narrow band emission near 2150 cm$^{-1}$ (at 733 nm). The key question whether this is the band due to light scattering or due to luminescence. One can wait probably the direct respond from the experiment with the change of excitation wavelength, but this attempt was not successful (see previous paragraph). Instead of answer, the experiment added one more fact to explain – the selectivity of excitation of the band mentioned above. This fact is necessary to take into considerations.

Evident possibility was that some semiconducting Cu-oxides are present on the surface and the band 2150 could originate from one of those oxides. But the $E_g$ values for those oxides (2,2eV for $Cu_2O$ and 1,4eV for CuO [2]) are far from energy position of interest (733 nm <-> 1,69 eV) .

The most essential fact, to our opinion, is absence of effect (the narrow band emission) from sample with smooth Cu substrate. We suppose that in any case – is it the light scattering or is it the luminescence, the surface enhancement (SE) plays the main role in ability to observe the band. Indeed, the plasmon resonance means local increase of the exciting light intensity in area of resonance. In it's turn it should increase intensity of any emission – of the luminescence and of the scattering as well. So, this fact in an argument in favor of SE involvement.

Observation of the bands in region 220-550 cm$^{-1}$ and the shift of their positions also shows that the involvement of the surface plasmons to explanation is very likely. These bands will be discussed in some other paper.

The small width of 2150 cm$^{-1}$ band is essential and demands separate explanation. In frames of plasmon hypothesis one of the main factors influences the line position is the separation between nanoparticle and substrate (or between tips of two metals). The oxide film on metal (metals) can serve as a regular spacer between them. Due to similarity over the surface, oxide film may provide the similar distance between and thus similar spectral position of resonance. As a result, the inhomogenous broadening may be small. The heating of sample in our case should influence the oxide film and to lead to disappearance of the band. At least that is true for Au: Au oxide can be formed in our conditions of sputtering for similar conditions and $Au_2O_3$ became unstable at temperatures above 200$^0$C [3-6].

Also the oxide film on Cu substrate can be changed by heating. Also our model explains why the effect can be observe with non-ideal surface of copper substrates, and can't be observed with smooth Cu surfaces. The non-ideal surface provides metallic tips, necessary for appearance of plasmon resonance, while smooth surface does not.

The homogeneity of oxide film can also be cause of selectivity of excitation.

Band near 1590 cm$^{-1}$, that can be seen in spectra of three-layer samples with interlayer $MoS_2$, never was published in papers concerning this material (e.g. see review [1]); the position of band is very close to position of graphite-like carbon band. Note, that it is not the case of graphen, because the last should demonstrate specific band near 2650 cm$^{-1}$ (see review [7]). One of the possibilities is that this band is due to traces of carbon, that gives intensive band in Raman spectra. The another ability – the band 1590 cm$^{-1}$ may appear due to higher order processes of light scattering in $MoS_2$; but now we have no arguments in favor of this hypothesis.

One more question is like that: can the band 2150 cm$^{-1}$ be due to some species absorbed at Cu-surface? We put the question to ourselves, and our experiments with molybdenum disulfide layer in samples were made in attempt to clarify this. The idea was to change a surface of copper before Au spattering. As the molybdenum disulfide can easily be sliced, we pilled it over copper. Of cause, we realized that surface was far from ideal in that case, but no doubt, the $MoS_2$ coverage should cause changes in surface. At the same time the band 2150 cm$^{-1}$ did not change it's position, at least the changes were not visible. So we consider that involvement of adsorbed substances in formation of spectra is unlikely.

One can assume that the gold particles from spattering and the $MoS_2$ particles from pilling lay on surface separately. The observation of $MoS_2$ multiple bands near 500 cm$^{-1}$ (fig.5) is in favor of such assumption. Nevertheless, due to chemical activity if sulfur, one can expect, that all surface was someway modified due to $MoS_2$ pilling. As we did not observe reaction from $MoS_2$ pilling on the position of line 2150 cm$^{-1}$, probably any other adsorbed species on surface did not influences much to the effect.

**Summery:**

A new narrow band was observed in emission spectra of Au film on Cu substrate. The band is not originates neither from substrate nor from Au film. The most probable explanation is that the band is due to inelastic light scattering at interface.

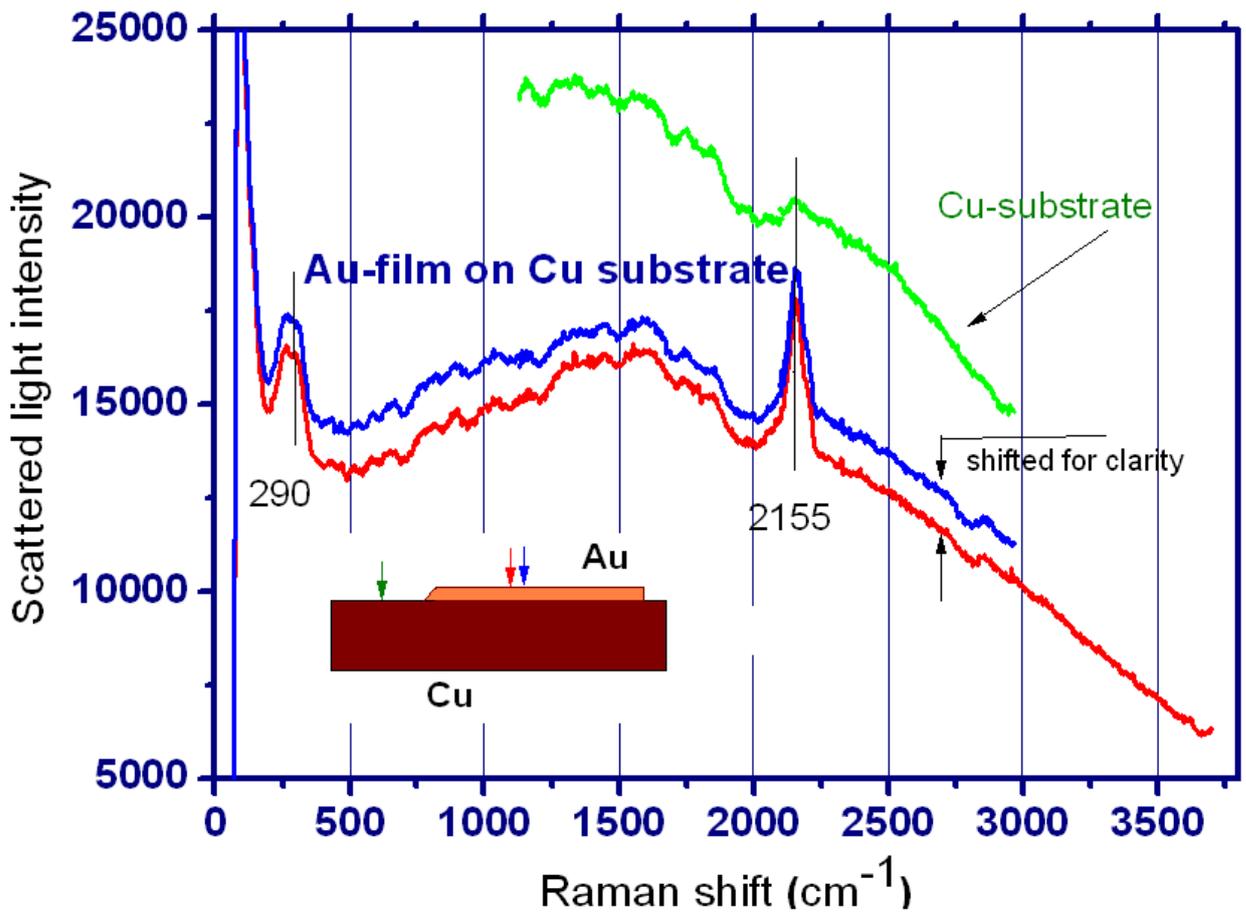

Fig 1. Emission spectra from Au-film on copper substrate. Red and blue spectra , two subsequently recorded spectra of the same point on of Au – film sample. The band near 290 cm$^{-1}$ will be discussed somewhere else. Green spectrum – was recorded from film-free part of copper substrate.

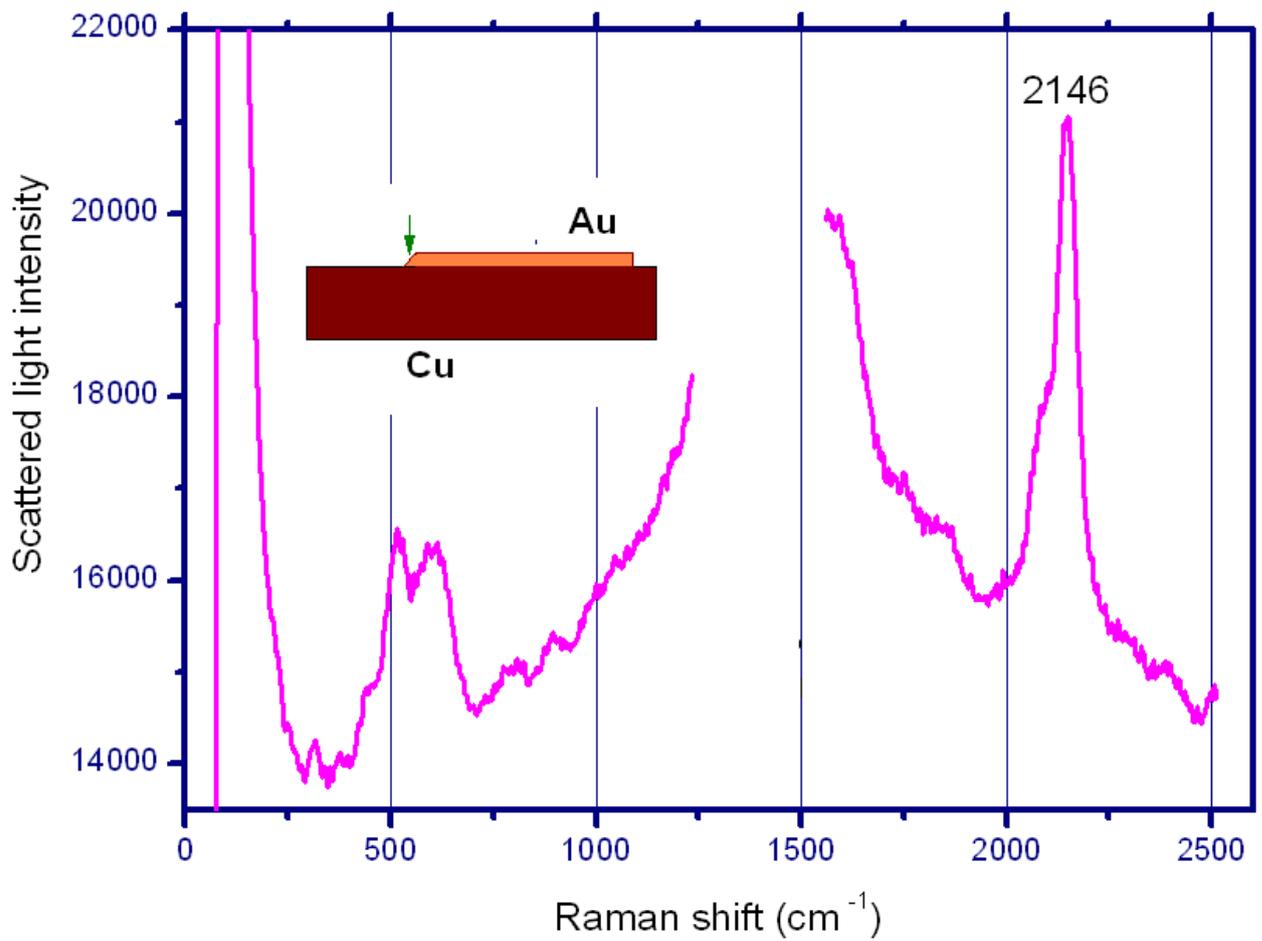

Fig.2 Similar to fig 1, but the excitation light is focused at the edge region of Au-film. One may see large shift of the position of low-energy band . Also the position of the Au-film line is changed about 10 cm$^{-1}$ .

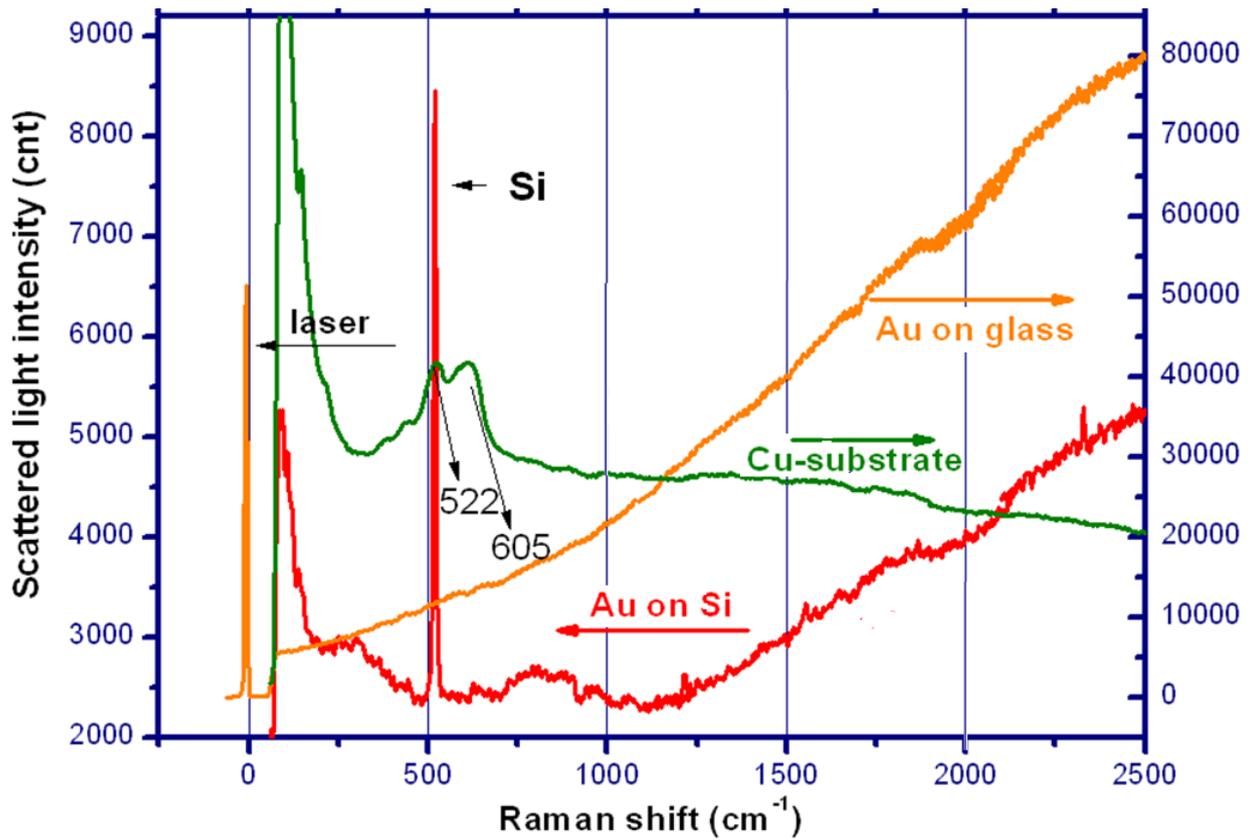

Fig.3 Raman spectra from the components of the sandwich samples

1. Red spectrum from Au-film on Si substrate. Easy to see well known band of silicon at 512 cm$^{-1}$ , but no any features can be observe near 2150 cm$^{-1}$ position.

2. Orange spectrum from Au film on glass

3 Green spectrum from Cu-substrate. Note well resolved band near 520-600cm$^{-1}$ , similar to band in spectra at fig.3.

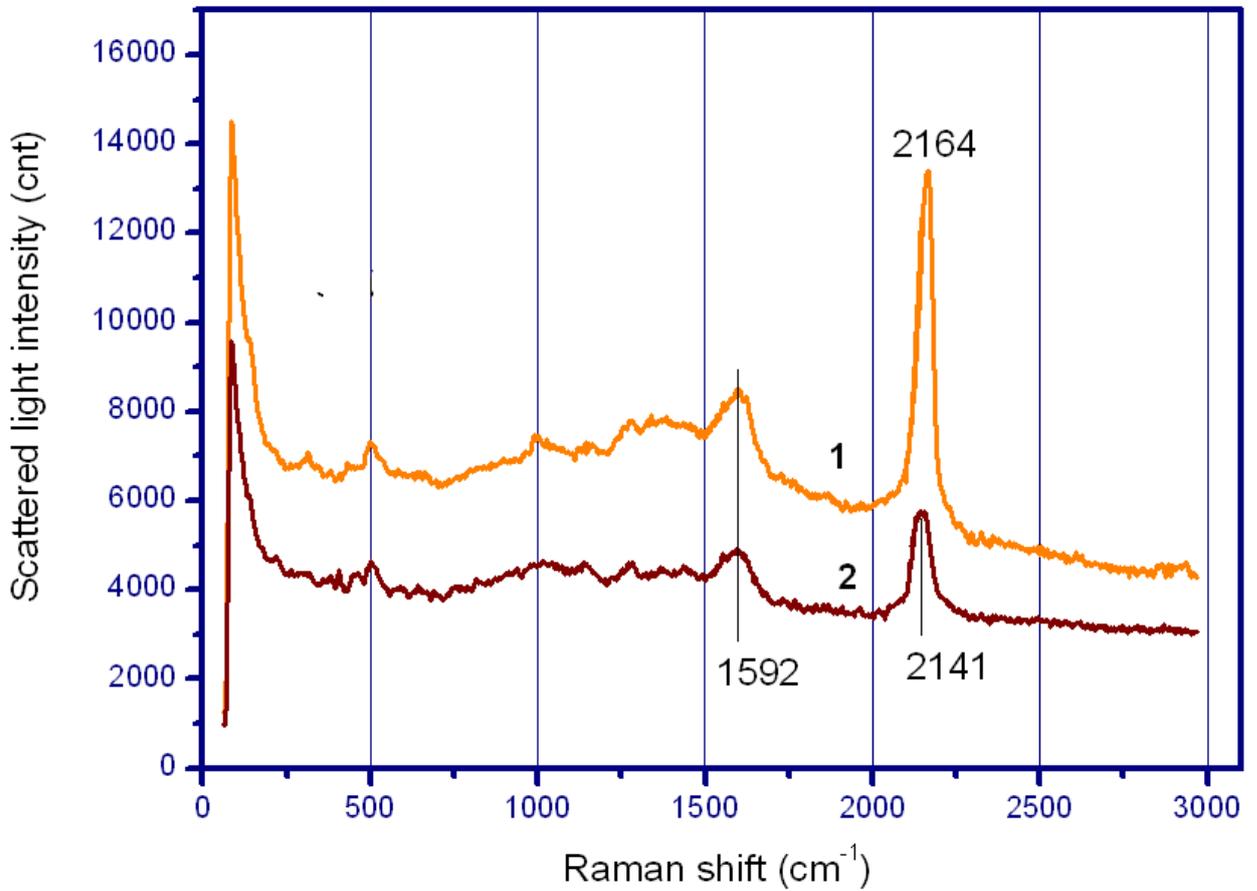

Fig 4  Emission spectra from tree-layer sample Au- $MoS_2$-Cu . The thickness of Au film at spectrum (1) is five times more, then Au thickness for film at spectrum (2) The weak traces of $MoS_2$ bands can be seen near 500 nm. Also band near 1590 cm$^{-1}$ is visible for tree-layer samples.

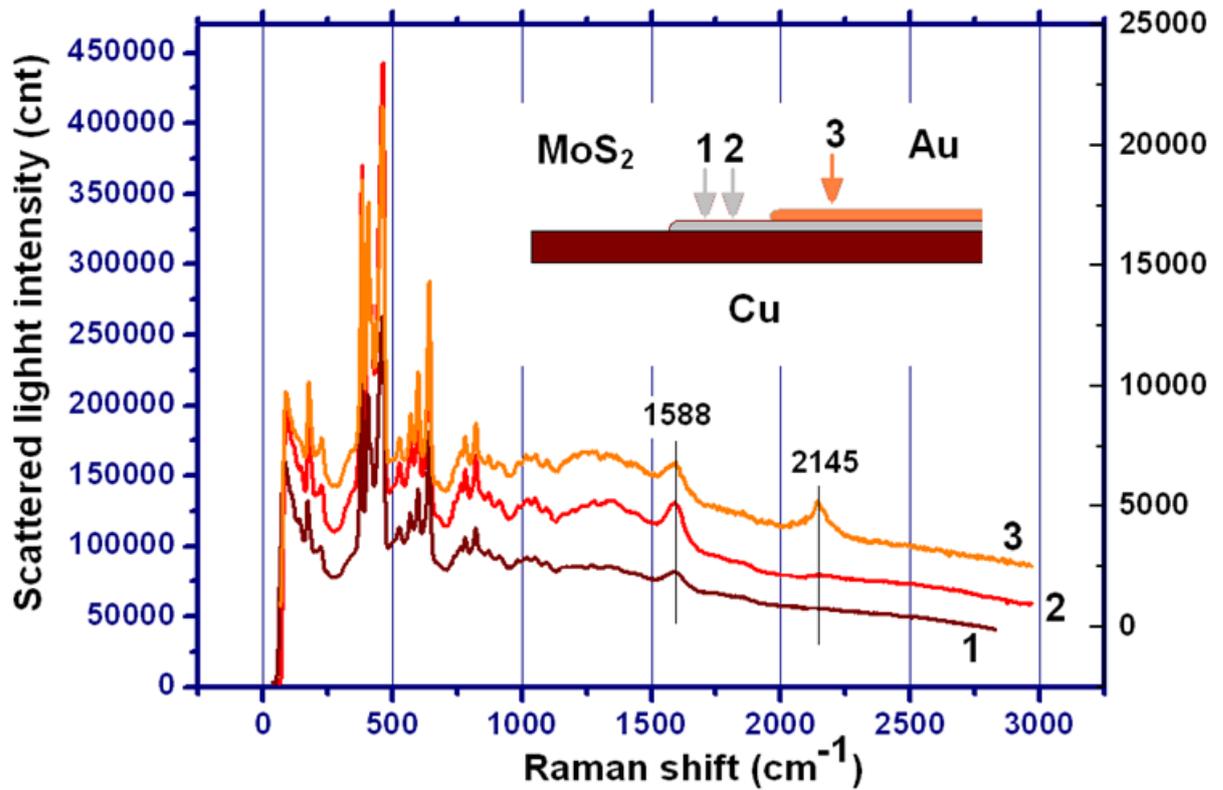

Fig.5 Spectra for tree-layer sample with thin Au film (spectrum 3) Spectra (1,2) were taken from Au-free edge of sample, where substrate was covered from Au-spattering. The multiple bands from $MoS_2$ can be seen around 500 cm$^{-1}$. The band near 1590 cm$^{-1}$ occurs in all spectra, but band 2150 cm$^{-1}$ is absent in spectra 1,2 .